# The Goldbeter-Koshland switch in the first-order region and its response to dynamic disorder

**Jianhua Xing[1, 2, 4], Jing Chen[3]**

1. Chemistry, Materials, and Life Sciences Directorate, University of California & Lawrence Livermore National Laboratory, Livermore, CA 94550

2. Department of Biological Sciences, Virginia Polytechnic Institute and State University, Blacksburg, VA, 24061-0406

3. Biophysics Graduate Program, University of California, Berkeley, Berkeley, CA 94720

4. Send correspondences to: jxing@vt.edu

# Abstract

In their classical work (Proc. Natl. Acad. Sci. USA, 1981, 78:6840-6844), Goldbeter and Koshland mathematically analyzed a reversible covalent modification system which is highly sensitive to the concentration of effectors. Its signal-response curve appears sigmoidal, constituting a biochemical switch. However, the switch behavior only emerges in the 'zero-order region', i.e. when the signal molecule concentration is much lower than that of the substrate it modifies. In this work we showed that the switching behavior can also occur under comparable concentrations of signals and substrates, provided that the signal molecules catalyze the modification reaction in cooperation. We also studied the effect of dynamic disorders on the proposed biochemical switch, in which the enzymatic reaction rates, instead of constant, appear as stochastic functions of time. We showed that the system is robust to dynamic disorder at bulk concentration. But if the dynamic disorder is quasi-static, large fluctuations of the switch response behavior may be observed at low concentrations. Such fluctuation is relevant to many biological functions. It can be reduced by either increasing the conformation interconversion rate of the protein, or correlating the enzymatic reaction rates in the network.

## Introduction

A biological system usually functions by regulating protein activities through protein interaction networks (PINs), which are built of interconnected modules. Some typical modules were summarized in a recent review by Tyson et. al.[1]. One of the most common modules in the PINs is the covalent modification system, which typically consists of a phosphorylated-dephosphorylated couple: $R \underset{\uparrow A}{\overset{\downarrow S}{\rightleftarrows}} R_P$ [2,3,4]. Here $R$ is the protein being covalently modified; its phosphorylated form, $R_P$, amounts to the response of the system. The kinase, $S$, enters as the external signal. $A$ denotes the phosphatase, which restores the protein to its 'non-response' form. This system shows "zero-order ultrasensitivity": sharp transitions occur in the signal-response curve when the modification enzymes, $S$ and $A$, are saturated by the substrate, $R$ and $R_p$, i.e. $[S]$ and $[A]$ much smaller than $[R]_t = [R] + [R_P]$. Eq.(1) gives the steady-state response of the system [4],

$$\text{module } R \underset{\uparrow A}{\overset{\downarrow S}{\rightleftarrows}} R_P : \qquad \frac{k_1[S]([R]_t - [R_P])}{K_{m1} + [R]_t - [R_P]} - \frac{k_2[A][R_P]}{K_{m2} + [R_P]} = 0 \qquad (1)$$

where $K_{m1}$ and $K_{m2}$ are the Michaelis-Menten constants. According to Eq.(1), $[R_P]$ appears as a sigmoidal, Goldbeter-Koshland function of the kinase concentration, $[S]$. The system thus behaves like a switch in response to external signals, constituting an important module in the PIN [5,6].

However, the assumption of saturated enzyme reaction does not always hold in real PINs. When $[S]$ and $[A]$ become comparable to $[R]_t$, the system transits from the zero-order into the first-order region, and the ultrasensitive switching behavior disappears in the simple two-component system. In the first part of this paper, we will show that the switching behavior in the first-order region can be restored by an additional cooperative mechanism of the phosphorylation reaction.

In the second part, we will discuss the effect of dynamic disorder on the biological switch. Dynamic disorder refers to the phenomenon that the 'rate constant' of a reaction appears as a stochastic function of time [7,8]. This phenomenon has attracted extensive experimental and theoretical studies since the pioneering work on ligand binding to myoglobin by Frauenfelder and coworkers [9]. Recently, single-molecule experiments directly showed the dynamic disorder in enzymatic reactions [10,11,12]. We refer the readers to some review articles and references therein for more information [13,14,15,16].

The dynamic disorder results from the relatively slow fluctuations of the protein conformation, either in the enzyme or the substrate. Conformation fluctuations result in fluctuations of the enzymatic reaction rate. In the traditional chemical reaction theories, it is assumed that the reaction occurs much more slowly than the conformational fluctuations; so the reaction rate observed on the slow reaction time scale appears as the ensemble average of the reaction rates of each conformation. But biochemistry and biophysics studies showed that the conformational fluctuations of proteins can be much slower than previously assumed, because of the rugged energy landscape of protein conformations [10,11,13,17,18,19]. The fluctuation time scale is sometimes comparable or even slower than that of the reaction itself. In this regime, the observed reaction rate becomes stochastic in time, reflecting the fluctuation of the enzyme conformations.

In this work, we will show that in the switch module studied in the first part, the dynamic disorder induced by the conformational fluctuations of the substrate protein mainly affects the variance of the system response, but not the ensemble average response. We also investigated two ways to reduce the noise originated by the dynamic disorder.

## Results

### *The sigmoidal switch outside the zero-order region*

The covalent modification system $R \underset{\uparrow A}{\overset{\downarrow S}{\rightleftarrows}} R_P$ only achieves zero-order ultrasensitivity when the substrate proteins are much more than the enzymes. Without this assumption, the intermediate products should not be reduced from the reaction pathway. The full pathway is presented in case a, Figure 1. The signal-response (SR) relationship of the system is given by the steady state solution of the governing equations (Eq.(2) and (3)):

$$\frac{d}{dt}\begin{pmatrix}[R]\\[RS]\\[R_P]\\[AR_P]\end{pmatrix}=\begin{pmatrix}-k_{1f}[S] & k_{1r} & 0 & k_2\\ k_{1f}[S] & -k_{1r}-k_1 & 0 & 0\\ 0 & k_1 & -k_{2f} & k_{2r}\\ 0 & 0 & k_{2f} & -k_{2r}-k_2\end{pmatrix}\begin{pmatrix}[R]\\[RS]\\[R_P]\\[AR_P]\end{pmatrix}=0$$

$$\frac{d}{dt}[S]=(k_{1r}+k_1)[RS]-k_{1f}[R][S]=0$$

(2)

with concentration constraints

$$[R]+[RS]+[R_P]+[AR_P]=[R]_t$$
$$[S]+[RS]=[S]_t$$

(3)

For mathematical simplicity, we assumed that the phosphatase is in such great excess that its concentration, $[A]$, remains approximately constant throughout time. So we absorbed it in $k_{2f}$. Similar treatment was made throughout the paper. Numerical calculations confirmed that relaxing this assumption does not qualitatively change the conclusion of this paper (not shown).

Experimentally, one can control as the external signal either $[S]_t$, the total substrate concentration, or $[S]$, the free substrate concentration (the $d/dt[S]$ equation is not needed in this case). Unfortunately the above system does not produce a desired sigmoidal SR curve with either signal form. A sigmoidal SR curve must have zero second derivative of the response to the signal at the inflection point. But Eq.(2) and (3) give identically negative second derivatives of $[R_P]$ to both $[S]$ and $[S]_t$ (relation (4), derived in Appendix S1, case a), which indicates the lack of sigmoidal behavior. The numerical result of this case is shown in Figure 2a.

$$\frac{d^2}{d[S]^2}[R_P]<0,\ \frac{d^2}{d[S]_t^2}[R_P]<0,\ \text{for}\ [S],\ [S]_t>0$$

(4)

Adding a tight-binding step to the phosphorylation reaction does not produce a sigmoidal curve, either. In case a2 (Figure 1), for example, $R$ and $S$ first form a weakly bound compound, $RS$. $RS$ then convert to the tightly bound form, $R*S$. Eventually $R*S$ proceeds to phosphorylation. In this system, the second derivative of the response to the signal is also mono-signed (see Appendix S1, case a2). Therefore no sigmoidal behavior emerges.

The above analysis suggests that nonlinear terms of $[S]$ are required to generate the sigmoidal behavior. We examined one such scheme (case b, Figure 1), inspired by the work of Sabouri-Ghomi *et. al.* [20]. In this case, we modified the model such that binding an additional *S* molecule to the intermediate compound, *RS*, greatly facilitates the phosphorylation ($k_1' \gg k_1$ in Eq.(5)). The governing equations become

$$\frac{d}{dt}\begin{pmatrix}[R]\\ [RS]\\ [R_P]\\ [AR_P]\\ [SRS]\end{pmatrix} = \begin{pmatrix}-k_{1f}[S] & k_{1r} & 0 & k_2 & 0\\ k_{1f}[S] & -k_{1r}-k_1-k_{1f}'[S] & 0 & 0 & k_{1r}'\\ 0 & k_1 & -k_{2f} & k_{2r} & k_1'\\ 0 & 0 & k_{2f} & -k_2-k_{2r} & 0\\ 0 & k_{1f}'[S] & 0 & 0 & -k_{1r}'-k_1'\end{pmatrix}\begin{pmatrix}[R]\\ [RS]\\ [R_P]\\ [AR_P]\\ [SRS]\end{pmatrix} = 0 \quad (5)$$

$$\frac{d}{dt}[S] = (k_{1r}+k_1)[RS] + (k_{1r}'+2k_1')[SRS] - k_{1f}[R][S] - k_{1f}'[RS][S] = 0$$

with concentration constraints

$$[R]+[RS]+[R_P]+[AR_P]+[SRS]=[R]_t,$$
$$[S]+[RS]+2[SRS]=[S]_t \quad (6)$$

The additional mechanism brings about a nonlinear term of $[S]$ (see Eq.(5)) and leads to the desired sigmoidal response (Figure 2a). Figure 2b shows the concentrations of the various protein forms as functions of the signal $[S]_t$. At low kinase concentration, most protein substrates are sequestered in the single-kinase intermediate, *RS*; in this regime, $[RS]$ increases steadily with $[S]_t$. Beyond a critical kinase concentration, though, the double-kinase intermediate, *SRS*, starts to form and stream through the phosphorylation reaction. This turnover results in the corresponding sudden drop in $[RS]$ and abrupt rise in the final signal, $[R_P]$. Our numerical analysis also revealed bistability in this case, which was confirmed by the analysis with the Chemical Reaction Network Toolbox [21]. This scheme of bistability is additional to that reviewed by Kholodenko [6]. Further analysis is necessary for the bistability.

For the same circuit, we also carried out the calculation when the control signal is the free kinase concentration, $[S]$ (reducing the $d/dt[S]$ equation). Our numerical analysis showed that the SR curve, i.e. the $[R_P]$-$[S]$ curve, is only mildly sigmoidal over a wide parameter range.

## *Effects of dynamic disorder*

In this part, we will investigate the effect of dynamic disorder on the sigmoidal switch, in particular, on the circuit presented in case b from the previous section. We adopted a multistate model studied by Kou et. al [22], in which the substrate protein assumes $N$ different conformations, $R_1, ..., R_N$. All the other forms of the protein possesses $N$ corresponding conformations, for example, $RS_1,..., RS_N$ for $RS$. Only matching conformations of the reactant and the product are admissible in chemical transitions, e.g. $R_{P2} \rightarrow AR_{P2}$, but not $R_{P2} \rightarrow AR_{P3}$. Some or all of the chemical reaction rates vary with different conformations. As the proteins randomly change their conformations, the average rates of the reactions undergo temporal fluctuations, or dynamic disorder. Since loops of reversible reactions exist in the system, e.g. $R_1 \leftrightarrow R_2 \leftrightarrow RS_2 \leftrightarrow RS_1 \leftrightarrow R_1$, the reaction rates on the loops have to satisfy the constraint of detailed balance. The products of rates in the two directions of the loop have to be equal. To save the trouble of maintaining this constraint, we only considered the dynamic disorder in the irreversible reactions, like $SRS \rightarrow R_P$, i.e. only $k_1, k_1', k_2$ are allowed to fluctuate along the conformational coordinate. Interconversion occurs between each pair of conformers of the same protein form. For simplicity, a uniform rate was used for all the conformation interconversion. The governing equations of the system consist of Eq.(5) repeated over the $N$ conformations, as well as the equations for conformation interconversion.

This model can be regarded as a discrete representation of the continuous, coupled diffusion-reaction model widely used in studies of protein motors and other macromolecules [23,24,25,26,27,28,29,30]. In the continuous model, the molecule of interest assumes several chemical states. At a given chemical state, the molecule diffuses along a conformational coordinate. Transitions between the chemical states happen vertically, i.e. without simultaneous displacement along the conformational coordinate. This is because the chemical transition is generally characterized as a barrier-crossing process: the waiting time for a transition (determined by the rate constants) may be long, but the actual transition happens on a very short time scale. This time scale usually allows no resolvable displacements along other degrees of freedom. In our model, the different complexes (*RS, SRS*, etc.) are analogous to the chemical states in the continuous model, and the conformational coordinate is discretized, accounting for the energy minima on the rugged energy landscape.

We first studied the dynamic disorder of $k_1^{'}$, since the facilitated phosphorylation, $SRS \rightarrow R_P$, is the key step that restores the desired switching behavior in case b (Figure 1b). We assumed that $k_1^{'}$ obeys the gamma distribution over the conformations (mathematical expression given in Methods). Figure 3 compares the resulting ensemble SR curve, $\overline{[R_P]}$ vs $\overline{[S]_t}$, with the SR curve of each conformation, $[R_{Pi}]$ vs $[S]_{ti}$. Here $\overline{[R_P]}$ and $\overline{[S]_t}$ are the ensemble concentration of the signal and the response, i.e. $\overline{[R_P]} = \sum_{i=1}^{N}[R_{Pi}]$, $\overline{[S]_t} = [S] + \sum_{i=1}^{N}\left([RS]_i + 2[SRS]_i\right)$. And $[S]_{ti}$ is the scaled signal concentration for the $i$-th conformation, $[S]_{ti} = [S] + N\left([RS]_i + 2[SRS]_i\right)$. In the limit of zero conformation interconversion rate, the SR curve of the $i$-th conformation corresponds to the scenario as if only this conformation exists.

Figure 3a shows that the ensemble SR curve is not significantly affected by the conformation interconversion rate, $r_{int}$. Therefore, the effect of dynamic disorder on the system is barely noticeable if the response is measured at bulk. But the variance of the response strongly depends on $r_{int}$. When the conformation interconversion is much slower than the chemical reactions, the reflection point of the SR curve shifts up to 20% of the original value (Figure 3b). This variance is dramatically diminished when $r_{int}$ increases to two orders of magnitude below the mean value of the disordered reaction rate (Figure 3c). Figure 4 summarizes the variances resulting from several trial cases. Dynamically disordered $k_2$ (case a) generates comparable variances as the disordered $k_1^{'}$ (case b) does, while disordered $k_1$ (case c) generates much smaller variances. This is because $k_2$ and $k_1^{'}$ are both associated with the main reaction pathway in the system, but $k_1$ is not (cf. Figure 1b). Case d of Figure 4 shows that perfect correlation between the disordered rates can also diminish the variance. In this case, $k_1$, $k_1^{'}$ and $k_2$ share exactly the same distribution over the conformations; the variance of the response reduces and becomes invariant to the interconversion rate, $r_{int}$. This happens because change of the kinase activity is perfectly compensated by the change of the phosphatase activity. However, this phenomenon is not robust at all. As shown in case e of Figure 4, broadening the distribution of $k_1^{'}$ immediately restores the large variance.

## Discussions

This work has two focuses. First we discussed the detailed mechanisms of the switch module in the first-order region. It is well known that the covalently-modification system studied by Goldbeter and Koshland cannot produce the switching behavior in this region. To recover this behavior, we grafted onto the Goldbeter-Koshland system the cooperative binding mechanism, which produces sigmoidal responses by itself. When the reaction is facilitated by the binding of two enzyme molecules, the system regains the remarkable Hill factor in the first-order region. While experimental studies on the proposed mechanism is lacking, it is common for enzymes to work in the form of dimers during signal transduction. Further experimental studies are necessary to test the theoretical result for the basic switch module here.

Next we discussed the effect of dynamic disorder on the switch module. The ensemble averaged behavior is nearly insensitive to dynamic disorder. But when the number of protein molecules is small, fluctuations of the module response will significantly affect the functioning of the switch. The critical signal level for the transition of response can shift up to 20% when the reaction rates fluctuate slowly. This study raises two important questions. First, has the nature evolved to reduce the effect of dynamic disorders in PINs, especially some vital ones that require high robustness? Our study suggests two possible mechanisms to reduce the noise: increasing protein conformation interconversion rates, or correlating the distributions of the chemical reaction rates in the PIN. Second, can the dynamic disorder in one module of the PIN actually reduce the overall noise of the PIN? Noise effect and reduction in a biological network is an actively studied field [31,32,33,34,35,36,37,38,39,40,41,42,43,44]. The dynamic disorder, as one source of noise, can have broad time scales, and interplay with other noises. Theses noises may offset each other in the overall behavior. For example, in this work only the substrate protein has different conformations. But in a real PIN with cascades of such phosphorylation-dephosphorylation cycles, the substrate of one cycle serves as the kinase or phosphatase of the next cycle. Novel behaviors will probably emerge with the complete set of dynamic disorders built in. This is a future direction to follow.

In this work we focused on the phosphorylation-dephosphorylation cycle. Same conclusions apply to other signal transduction modules with similar kinetic structures. One example is the GTPase cycle [45,46,47]. A membrane-bound GTPase switches between a GTP-bound form, and a GDP-bound form, which assume different enzyme activity. Such a protein can detect GTP concentration as the signal, and respond with effects on the downstream biochemical pathways. The dynamics of such GTPase cycle is mathematically isomorphic to that of the phosphorylation-dephosphorylation cycle [47]. Thus, it should demonstrate the same switching behavior and effect of the dynamic disorder.

## Methods

In the first part, we used both the downhill simplex method and the simulated annealing method to search the parameter space of rate constants for the minimum of the following function [48]:

$$\int_0^1 \left| [R_P(S)] - T(S) \right| dS \approx \frac{1}{M} \sum_{i=1}^{M} \left| [R_P(S_i)] - T(S_i) \right| \tag{7}$$

Here $[R_P(S_i)]$ is the calculated $R_P$ concentration as a function of the signal strength, and $T(S)$ is the desired sigmoidal SR curve. In all the calculations $T(S)$ assumes the 'Goldbeter-Koshland' function form,

$$G(u,v,J,K) = \frac{2uK}{v - u + vJ + uK + \sqrt{(v - u + vJ + uK)^2 - 4(v-u)uK}} \tag{8}$$

with $u = 2S, v = 1, J = K = 0.05$.

In the second part, the dynamically disordered enzymatic reaction rates were computed from a gamma distribution with mean $ab$ and variance $ab^2$ (Eq.(9)). $a$ and $b$ are given in the corresponding figure captions.

$$p(k;a,b) = \frac{k^{a-1}e^{-k/b}}{b^a \Gamma(a)} \tag{9}$$

The above continuous distribution is discretized in the following way. First the parameter coordinate was divided into *N* bins with equal accumulated probability, i.e.

$$\int_{\xi_{i-1}}^{\xi_i} p(k)dk = 1/N, i = 1,\ldots,N$$

where $\xi_i$'s are the boundaries of the bins, with $\xi_0 = -\infty, \xi_N = \infty$. Then the *N* discretized rate constants were chosen as,

$$k_i = \frac{\int_{\xi_{i-1}}^{\xi_i} dk\, p(k)k}{\int_{\xi_{i-1}}^{\xi_i} dk\, p(k)} = N\int_{\xi_{i-1}}^{\xi_i} dk\, p(k)k, i = 1,\ldots,N$$

## Acknowledgements

We thank Professor Hong Qian, Professor John Tyson, and Mr. Wei Min for reading the manuscript and providing many helpful suggestions. We also thank the anonymous reviewer for many helpful suggestions, e.g, the same discussion may apply to the GTPase cycle.

# Figures

**Figure 1** Candidate schemes for the sigmoidal switch. Case a (upper panel within the shaded box): the full pathway of the zero-order ultrasensitive switch studied by Goldbeter and Koshland. The protein $R$ is phosphorylated and dephosphorylated by enzyme $S$ and $A$ respectively. The phosphorylated form $R_P$ represents the response and the enzyme $S$ the signal. Case a2 (the entire upper panel): $R$ and $S$ form a loosely bound complex, then a tightly bound one preceding the phosphorylation step. Case b (lower panel, the shaded area is the same as case a): $RS$ can bind another $S$ molecule, which accelerate the phosphorylation reaction $k_1$.

**Figure 2** SR curves of the candidate schemes. (a) Example SR curves obtained with the schemes shown in Figure 1. The G-K function is the result Goldbeter and Koshland obtained for the zero-order region. Parameters of other cases are obtained by fitting the steady state solutions to the G-K function (see Methods), and are listed in Table 1. (b) Concentrations of various protein forms as a function of the total signal concentration in case b, the only scheme showing the switch behavior.

**Figure 3** SR curves of case b in the presence of dynamic disorder in $k_1'$. A set of $k_1'$ of 10 different conformations are computed from a gamma distribution $p(k) = [1/(b^a \Gamma(a))] k^{a-1} \exp(-k/b)$ (see Methods). For the results shown here, $a = 4$, $b = 10/a$, so that the mean rate constant is 10, the $k_1'$ value used in the first section (Table 1). (a) The ensemble averaged response curves with the conformation $k_1'$ interconversion rate $r_{int} = 10^{-5}$ (solid line) and $r_{int} = 10^{-1}$ (dashed line). (b) Responses by individual conformers with $r_{int} = 10^{-5}$ (dotted lines). The ensemble averaged response is also shown in comparison as the solid line. (c) Same as b with $r_{int} = 10^{-1}$.

**Figure 4** Variances of the SR curves change with the conformation interconversion rate, $r_{int}$, with different sets of disordered parameters. Upper: $r_{int}$ vs. relative variances of the critical signal level. The critical signal level is defined as $[S]_{ti}$ at half the plateau responses (i.e. $[R_P]_i = [R_P]_i([S]_t \to \infty)/2$).
Bottom: $r_{int}$ vs. relative variances of the plateau response (i.e. $[R_P]_i([S]_t \to \infty)$). Different sets of disordered parameters denoted in the legend: a) $k_2$ is computed from the gamma distribution with $a = 4, b_2 = 1/a$; b) $k_1'$ is computed from the gamma distribution with $a = 4, b_1' = 10/a$; c) $k_1$ is computed from the gamma distribution with $a = 4, b_1 = 0.008/a$; d) all the enzymatic reaction rates, $k_1$, $k_1'$ and $k_2$, are computed from the gamma distribution with $(a, b_1; a, b_1'; a, b_2)$; e) and $k_2$, come from the gamma distribution with $(a, b_1; a, b_2)$, but $k_1'$ comes from the gamma distribution with $(a/2, 2b_1')$. The parameters of the gamma distributions in case a, b, c are chosen so that the mean rates equal the ones used in absence of the dynamic disorder. Note that the parameter $a$, which determines the width of the distribution, were chosen the same except in the last case for $k_1'$.

## Tables

Table 1 Model parameters.

|  | $k_1$ | $k_{1f}$ | $k_{1r}$ | $k_{1f}'$ | $k_{1r}'$ |  | $k_2$ | $k_{2f}$ | $k_{2r}$ | $[R]_t$ |
|---|---|---|---|---|---|---|---|---|---|---|
| Case a | 5 | 0.06 | 20 |  |  |  | 1 | 0.06 | 20 | 1 |
| Case b | 0.006 | 400 | 12 | 40 | 670 | 10 | 1 | 4 | 79 | 1 |

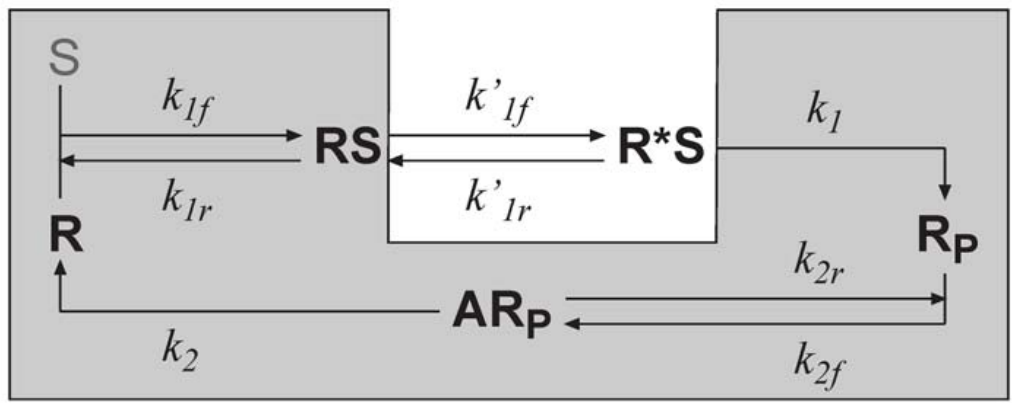

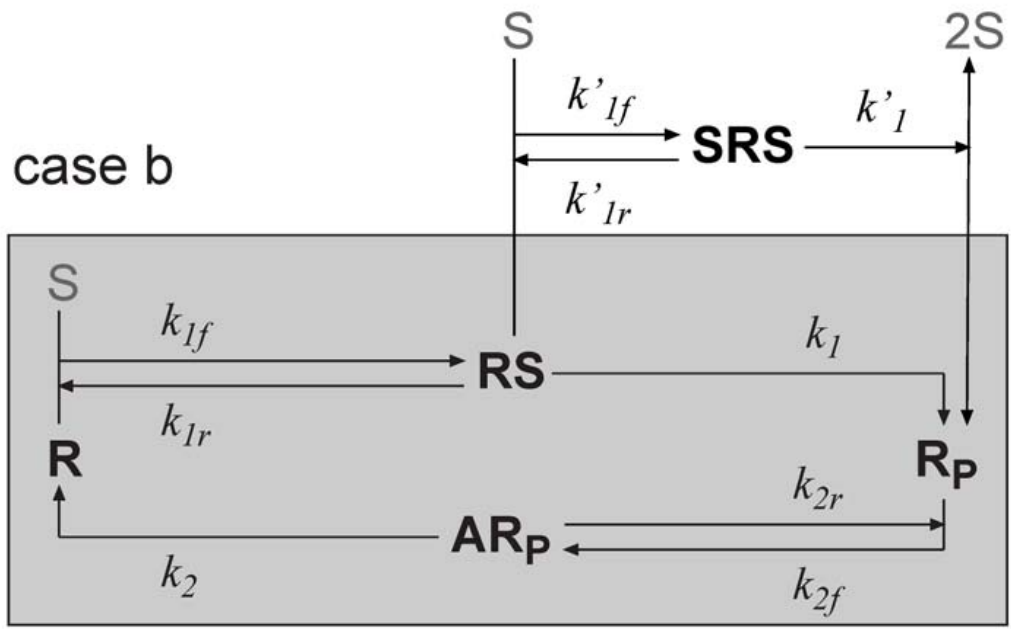

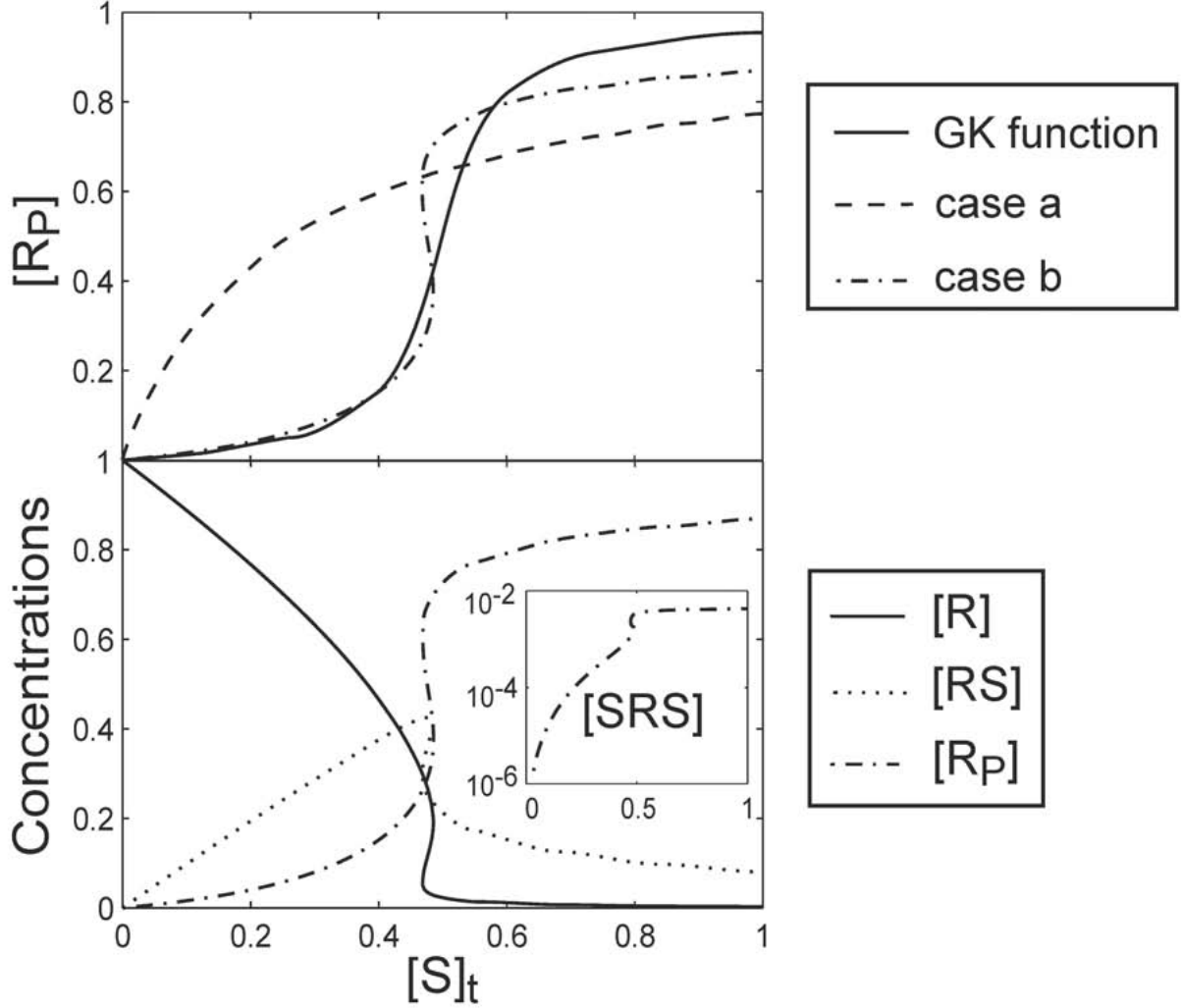

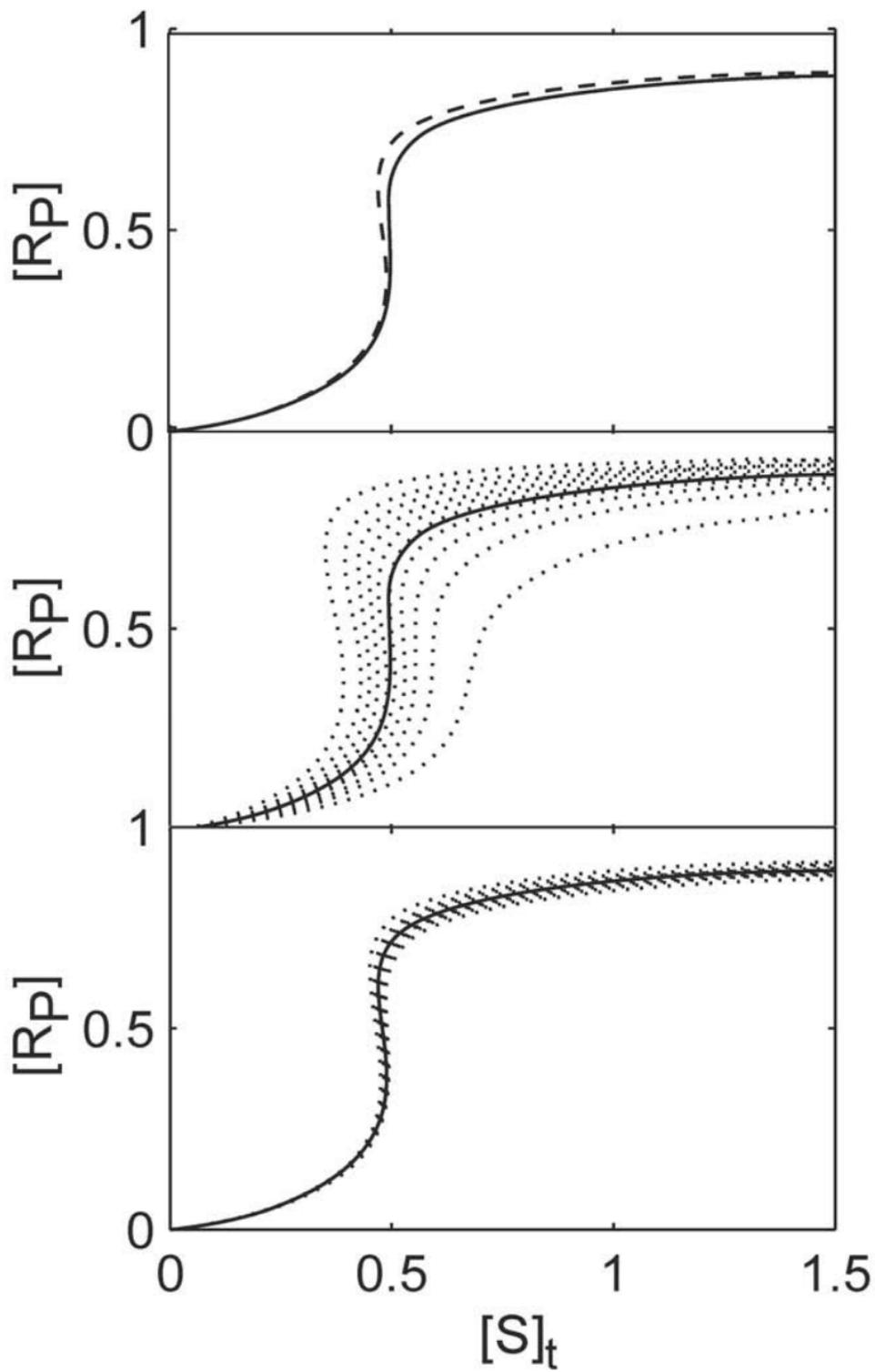

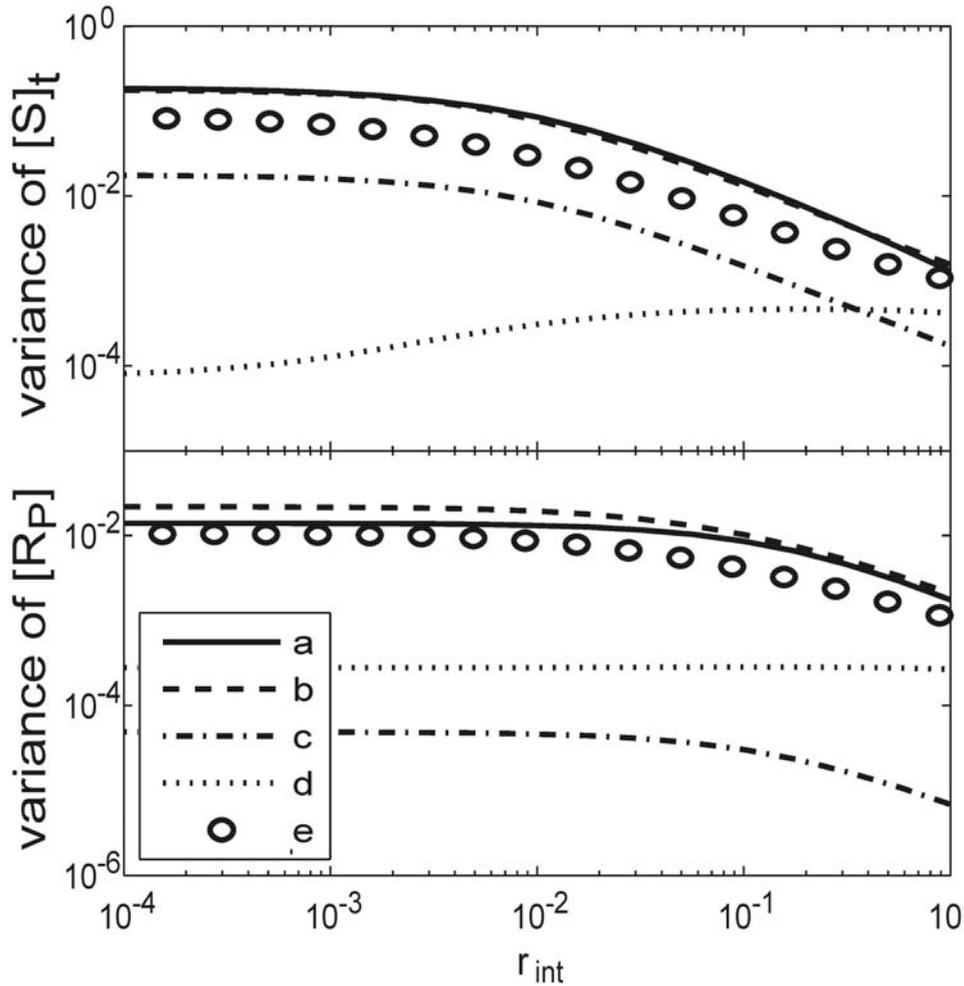

# Appendix S1

Here we show that case a and a2 cannot generate sigmoidal behavior, because the second derivative of $[R_P]$ as function of $[S]$ or $[S]_t$ is always monotonic.

## Case a

We examined the steady-state solution of Eq. (2) and (3) in the main text. First other quantities can be expressed by $[R]$,

$$[RS] = \frac{[S]_t [R]}{K_{m1} + [R]}$$

$$[AR_P] = \frac{k_1}{k_2}[RS] = \frac{k_1}{k_2} \frac{[S]_t [R]}{\left(K_{m1} + [R]\right)} \tag{S1}$$

$$[R_P] = K_{m2}[AR_P] = \frac{k_1}{k_2} \frac{K_{m2}[S]_t [R]}{\left(K_{m1} + [R]\right)}$$

Then the physically meaningful solution of $[R]$ is

$$[R] = \frac{-B + \sqrt{B^2 + 4[R]_t K_{m1}}}{2}$$

$$[R_P] = \frac{k_2 K_{m2}\left(a[S]_t + K_{m1} + [R]_t - \sqrt{a^2[S]_t^2 + 2ab[S]_t + \left([R]_t + K_{m1}\right)^2}\right)}{2a} \tag{S2}$$

where,

$$K_{m1} = \frac{k_{1r} + k_1}{k_{1f}}, K_{m2} = \frac{k_{2r} + k_2}{k_{2f}},$$

$$B = \left(1 + k_2 + k_2 K_{m2}\right)[S]_t + K_{m1} - [R]_t = a[S]_t + b$$

Then,

$$\frac{2a}{k_2 K_{m2}} \frac{d}{d[S]_t}[R_P] = a - \frac{a^2[S]_t + ab}{\sqrt{a^2[S]_t^2 + 2ab[S]_t + ([R]_t + K_{m1})^2}}$$

$$\frac{2}{ak_2 K_{m2}} \frac{d^2}{d[S]_t^2}[R_P] = \frac{-4[R]_t K_{m1}}{\sqrt{a^2[S]_t^2 + 2ab[S]_t + ([R]_t + K_{m1})^2}} < 0$$

The second derivative of $[R_P]$ to $[S]_t$ is monotonic, and approaches zero only when $[S]_t \to \infty$. Therefore, SR curve is hyperbolic and shows no sigmoidal behavior.

Next we examine the function dependence of $[R_P]$ on $[S]$, the concentration of free $S$ molecules. The steady state solution of $[R_P]$ has the form,

$$[R_P] = \frac{[S] + \alpha}{\beta[S] + \gamma} \tag{S3}$$

One can easily show that its second derivative to $[S]$ is mono-signed. Therefore, case a does not have a sigmoidal behavior whether the total concentration or free concentration of $S$ is controlled as the signal.

## Case a2

The governing equations of this case (Figure 1a) read

$$\frac{d}{dt}\begin{pmatrix}[R]\\ [RS]\\ [R^*S]\\ [R_P]\\ [AR_P]\end{pmatrix} = \begin{pmatrix}-k_{1f}[S] & k_{1r} & 0 & 0 & k_2\\ k_{1f}[S] & -k_{1r}-k_{1f}^{"} & k_{1r}^{"} & 0 & 0\\ 0 & k_{1f}^{"} & -k_{1r}^{"}-k_1 & 0 & 0\\ 0 & 0 & k_1 & -k_{2f} & k_{2r}\\ 0 & 0 & 0 & k_{2f} & -k_{2r}-k_2\end{pmatrix}\begin{pmatrix}[R]\\ [RS]\\ [R^*S]\\ [R_P]\\ [AR_P]\end{pmatrix} = 0 \tag{S4}$$

$$\frac{d}{dt}[S] = (k_{1r} + k_1)[RS] - k_{1f}[R][S] = 0$$

with concentration constraints

$$[R]+[RS]+[R^*S]+[R_P]+[AR_P]=[R]_t$$
$$[S]+[RS]+[R^*S]=[S]_t \tag{S5}$$

In this case, an additional intermediate step is added upon case a. After some tedious but straightforward derivation, one obtains

$$[R]=\frac{-B+\sqrt{B^2+4AC}}{2A}$$

$$[R_P]=\frac{\frac{k_1}{2Ak_2}K_{m2}\left(a[S]_t - \sqrt{a^2[S]_t^2 + 2ab[S]_t + b^2 + 4AC} + 2A[R]_t + ab\right)}{\left(\frac{k_1}{k_2}+\frac{k_1}{k_2}K_{m2}+1+K^{"}_{m1}\right)}$$

with

$$K_{m1}=\frac{k_{1r}}{k_{1f}}, K_{m2}=\frac{k_{2r}+k_2}{k_{2f}}, K^{"}_{m1}=\frac{k^{"}_{1r}+k_1}{k^{"}_{1f}},$$

$$A=1+K^{"}_{m1}$$

$$B=k_{1f}[S]_t\frac{k_1}{k_2}\left(1+K_{m2}+\frac{k_2}{k_1}+\frac{k_2}{k_1}K^{"}_{M1}\right)+k_{1r}K^{"}_{m1}+k_1-\left(1+K^{"}_{m1}\right)[R]_t=a[S]_t+b$$

$$C=\left(k_{1r}K^{"}_{m1}+k_1\right)[R]_t$$

Based on these equations it can be proved mathematically that the second derivatives of $[R_P]$ to $[S]$ and $[S]_t$ are both mono-signed. Therefore, case a2 gives no sigmoidal behavior.